\documentclass[final,twocolumn,authoryear]{elsarticle}

\usepackage[authoryear]{natbib}
\biboptions{authoryear}

\usepackage{amsmath}
\usepackage{amssymb}
\usepackage{graphicx}
\usepackage{booktabs}
\usepackage{siunitx}

\usepackage{caption}
\usepackage{subcaption}

\usepackage{tabularx}
\usepackage{array}
\usepackage{microtype}
\usepackage{hyperref}

\journal{Advances in Space Research}

\begin{document}

\begin{frontmatter}

\title{Physicochemical properties of lunar regolith simulant for in situ oxygen production}

\author[inst1,inst2]{Alyssa Ang De Guzman}
\author[inst2,inst4]{Anish Mathai Varghese}
\author[inst1]{Saif Alshalloudi}
\author[inst2,inst3]{Lance Kosca}
\author[inst2,inst3]{Kyriaki Polychronopoulou}
\author[inst1,inst2]{Marko Gacesa\corref{cor1}}

\cortext[cor1]{Corresponding author}
\ead{marko.gacesa@ku.ac.ae}

\affiliation[inst1]{
  organization={Department of Physics, Khalifa University},
  city={Abu Dhabi},
  postalcode={P.O. Box 127788},
  country={United Arab Emirates}
}

\affiliation[inst2]{
  organization={Center for Catalysis and Separation (CeCaS), Khalifa University},
  city={Abu Dhabi},
  postalcode={P.O. Box 127788},
  country={United Arab Emirates}
}

\affiliation[inst3]{
  organization={Department of Mechanical and Nuclear Engineering, Khalifa University},
  city={Abu Dhabi},
  postalcode={P.O. Box 127788},
  country={United Arab Emirates}
}

\affiliation[inst4]{
  organization={Department of Chemical and Petroleum Engineering, Khalifa University},
  city={Abu Dhabi},
  postalcode={P.O. Box 127788},
  country={United Arab Emirates}
}

\begin{abstract}
Permanent lunar settlements will rely on in situ oxygen production from regolith for life support and propulsion. While oxygen is abundant in lunar materials, it is chemically bound within metal oxides whose extractability depends strongly on regolith composition and processing strategy. In this study, we validate and characterize high-fidelity lunar regolith simulants representative of the lunar highlands and south pole using scanning electron microscopy with energy-dispersive X-ray spectroscopy, X-ray diffraction, Brunauer-Emmett-Teller surface area and pore structure analysis, and hydrogen temperature-programmed reduction. The simulants exhibit strong mineralogical and compositional fidelity to returned Apollo and Chang’e samples, with ilmenite confirmed as the most readily reducible oxygen-bearing phase. However, despite low ilmenite abundance, bulk highland simulants display favorable reduction behavior arising from distributed Fe-bearing silicate and glassy phases, as well as surface and microstructural properties that influence gas-solid interactions. Adsorption experiments with gases (H$_2$, CH$_4$, and CO$_2$) and water indicate that mineralogical heterogeneity and pore accessibility influence their uptake in simulants. These results indicate that oxygen extraction behavior in realistic lunar regolith is governed by whole-regolith response rather than ilmenite content alone, supporting the option of whole-regolith processing strategies for oxygen production in lunar in situ resource utilization architectures.
\end{abstract}


\end{frontmatter}

\section{Introduction}

As global interest in establishing sustained lunar presence increases, the development of robust in situ resource utilization (ISRU) strategies becomes essential. Among all in situ resources, oxygen is of primary importance, both for life support and as an oxidizer for propulsion systems. The Moon contains abundant oxygen, approximately 45 wt\% in returned Apollo samples, primarily bound within minerals such as ilmenite, olivine, pyroxene, and plagioclase \cite{sibille_state---art_2020}. However, oxygen is not readily accessible and must be liberated through energy-intensive chemical or electrochemical processes \cite{d_green_-situ_2019}. Consequently, the feasibility of oxygen production on the Moon depends strongly on regolith composition, mineralogy, and the selection of appropriate extraction pathways.

Lunar regolith is commonly classified into mare and highland types, reflecting the Moon’s major geological terrains. Mare regolith is basaltic and relatively enriched in ilmenite, making it well suited for hydrogen reduction-based oxygen extraction approaches \cite{sargeant_hydrogen_2020}. In contrast, highland regolith, dominant across much of the lunar surface and particularly relevant to polar regions, is feldspathic, anorthosite-rich, and typically contains very low ilmenite abundances \cite{slabic_lunar_2024,mckay1991lunar}. As a result, oxygen extraction strategies that rely exclusively on ilmenite reduction may be poorly suited for highland and polar deployments, motivating increased interest in alternative approaches such as molten regolith electrolysis, plasma-assisted extraction, and whole-regolith processing schemes \cite{schreiner_molten_2015}.
Recent analyses of returned Apollo and Chang’e-5 samples have further highlighted the mineralogical complexity and heterogeneity of lunar regolith. Apollo highland soils are dominated by plagioclase and commonly contain less than 2 vol\% ilmenite even in fine fractions \cite{taylor_mineralogical_2010}, while Chang’e-5 samples, though collected from a mare region, include highland clasts rich in calcic plagioclase and magnesian mafic minerals \cite{wang_first_2022}. These findings underscore that oxygen extraction systems must contend with mixed mineral assemblages rather than idealized, ilmenite-rich feedstocks. In response, a new generation of high-fidelity lunar regolith simulants has been developed to better reproduce the mineralogical and physicochemical properties of specific lunar terrains \cite{zou_development_2024,sun_developing_2017}. Earlier simulants were often discontinued, lacked sufficient fidelity, or were unsuitable for advanced ISRU testing, particularly in the context of oxygen extraction and additive manufacturing \cite{isachenkov_characterization_2022}.

Beyond serving as geological analogs, high-fidelity simulants provide a practical platform for evaluating how regolith composition, microstructure, and surface properties influence gas-solid interactions, reducibility, and volatile handling under ISRU-relevant conditions. In this context, the key question is not only whether a simulant matches lunar composition, but what it reveals about feasible oxygen extraction strategies for regions where ilmenite is scarce.

In this study, we validate and characterize the lunar highland simulants LHS-1 and LHS-2, as well as the lunar south pole simulant LSP-2, using SEM-EDX, XRD, BET surface area and pore structure analysis, and hydrogen temperature-programmed reduction (TPR). By combining chemical, structural, and textural characterization with reduction and adsorption measurements, we assess the extent to which whole-regolith processing can support oxygen extraction behavior despite low ilmenite content. Our results indicate that bulk highland simulants, particularly LHS-2, exhibit favorable gas-solid interaction and reduction behavior arising from distributed Fe-bearing silicate and glassy phases, rather than ilmenite alone. These findings inform the selection and optimization of oxygen extraction pathways for highland and polar ISRU architectures, where feedstock realism and system robustness are critical.

\section{Materials and Methods}
%\FloatBarrier

This study utilizes high-fidelity lunar regolith simulants produced by Exolith to accurately represent distinct soil types from the south pole (LSP-2) and highlands (LHS-1 and LHS-2) regions of the Moon  \cite{easter_effect_2024,long-fox_geomechanical_2023}. In this study, all references to LHS-1, LHS-2, and LSP-2 refer to this manufacturer’s product.

\begin{table}[b]
\centering
\caption{Reported simulant mineral compositions.}
\label{tab:reported_minerals}
\begin{tabular}{lcc}
\toprule
Component &
\begin{tabular}[c]{@{}c@{}}LHS-1 \& LHS-2\\(wt\%)\end{tabular} &
\begin{tabular}[c]{@{}c@{}}LSP-2\\(wt\%)\end{tabular} \\
\midrule
Anorthosite        & 74.4 & 90.0 \\
Glass-rich Basalt  & 24.7 & 10.0 \\
Ilmenite           & 0.4  & --   \\
Olivine            & 0.2  & --   \\
Pyroxene           & 0.3  & --   \\
\bottomrule
\end{tabular}
\end{table}

The phase compositions of the simulants used are reported in Table~\ref{tab:reported_minerals} \cite{SRT_LHS1_2023,SRT_LHS2_2023,SRT_LSP2_2023}. It should be noted that LHS-1 and LHS-2 are compositionally similar. The simulants differ in particle size: LHS-1 averages 100 µm, whereas LHS-2 is 60 µm. Pyroxene may be referred to as bronzite in Exolith specifications.

To characterize the simulants, scanning electron microscopy with energy dispersive X-ray spectroscopy (SEM-EDX) was used to examine particle size, morphology, and surface topology. This was conducted with the \textit{ThermoFisher} Phenom-XL Desktop SEM with an integrated EDX detector.

\begin{figure*}[t]
\centering
\begin{subfigure}{0.48\textwidth}
  \centering
  \includegraphics[width=\linewidth]{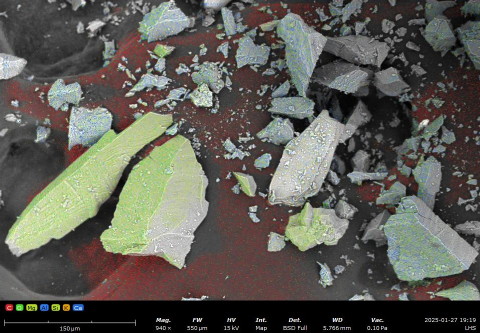}
  \label{fig:lhs2_edx_sub}
\end{subfigure}
\hfill
\begin{subfigure}{0.48\textwidth}
  \centering
  \includegraphics[width=\linewidth]{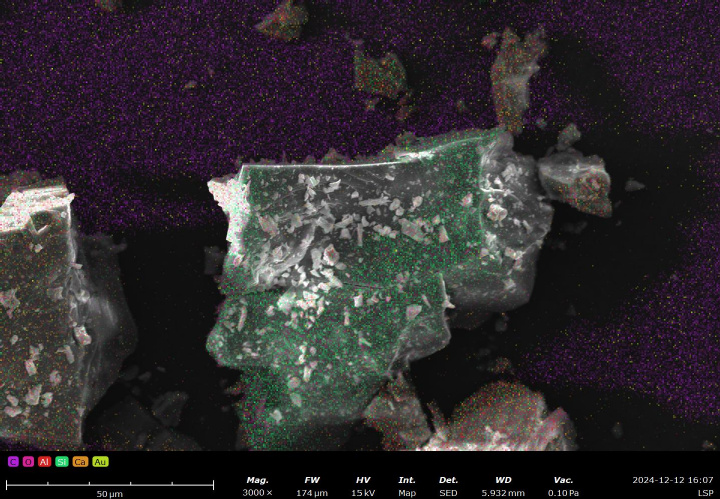}
  \label{fig:lsp2_edx_sub}
\end{subfigure}
\caption{EDX elemental maps of lunar regolith simulants: (a) LHS-2 (left) and (b) LSP-2 (right).}
\label{fig:edx_maps}
\end{figure*}

As shown in Figure~\ref{fig:edx_maps}, SEM-EDX mapping was used to visualize elemental distributions as color-mapped signal intensities. Carbon and gold signals are artifacts arising from the conductive mounting tape and sputter-coating process, respectively, and are not intrinsic to the simulant material. Excluding these artifacts, EDX mapping identifies six major elements within the simulant grains: O, Si, K, Mg, Al, and Ca.

The SEM-EDX maps reveal pronounced mineralogical heterogeneity at the grain scale, with oxygen-rich silicate and glassy matrices surrounding discrete Mg- and Al-rich grains. Large, compositionally continuous ilmenite domains are not observed at this scale, indicating that Fe- and Ti-bearing phases, where present, are likely fine-grained and spatially dispersed. The predominantly angular grain morphology and the presence of intergranular voids, rather than internal grain porosity, are qualitatively consistent with the low BET surface area and mesopore-scale pore sizes measured for the bulk simulants. This heterogeneous and dispersed phase distribution further aligns with the broad, multi-step reduction behavior observed during hydrogen TPR, discussed in Section 3.2.

To further verify mineralogical composition, particularly for phases that may be present below SEM-EDX detection limits, powder X-ray diffraction (XRD) analysis was conducted on untreated samples of LHS-1, LHS-2, and LSP-2 simulants, alongside their constituent components, using a \textit{Malvern Panalytical} diffractometer configured for high-resolution phase identification.

XRD results were recorded over a $2\theta$ range of $5^\circ$ to $85^\circ$ on the as-received powders. Raw diffraction profiles were processed using \textit{X'Pert HighScore}, employing background subtraction and K-$\alpha_2$ stripping algorithms prior to peak fitting, mineral phase identification, and comparative analysis across simulant compositions. Apollo~16 sample~\#64501 (from Ref.~\cite{taylor_modal_2019}) was employed as a benchmark for highland regolith composition in both LHS-1 and LHS-2. All patterns were corrected for zero-point displacement by applying a constant $2\theta$ offset to account for instrument drift during peak comparison \cite{hulbert_specimen-displacement_2023}.

Upon verifying the fidelity of the simulants, Brunauer--Emmett--Teller (BET) analysis was used to assess surface area properties of LHS-2 and its constituent minerals. A \textit{Micromeritics 3Flex} Surface Analyzer was used to perform N$_2$ adsorption--desorption at 77~K. Prior to measurement, samples were activated for 12~hours at 200~$^\circ$C under high vacuum. LHS-1 was not included in BET analysis as it is compositionally similar to LHS-2. The latter’s smaller particle size already lends to greater surface area exposure, corresponding to enhanced gas adsorption. Furthermore, gas--solid interaction experiments with CO$_2$, N$_2$, H$_2$, and CH$_4$ were conducted at 25$^\circ$C and up to 1~bar, as well as water adsorption isotherms of the materials measured at 25$^\circ$C and 100\% RH.

Finally, temperature-programmed reduction (TPR) using a \textit{Micromeritics AutoChem} was performed to characterize chemical interactions between solids and reacting gases, quantifying the strength of solid-oxygen interactions. In hydrogen TPR samples were pretreated under flowing Ar for 1 hour, starting from ambient to a peak temperature of 300 °C reached in 10 minutes. The samples consisted of two simulant mixtures (LHS-2, LSP-2) and their respective phases (ilmenite and olivine). The samples were all reduced at 900 °C. Moreover, in the context of oxygen extraction, it should be noted that ilmenite reduction is consistently identified as one of the most efficient and straightforward methods for oxygen extraction \cite{sargeant_hydrogen_2020,ramachandran_hydrogen_2006}, and thus TPR results for these ilmenite-containing simulants are expected to produce a similar finding.

\section{Results and Discussion}
%\FloatBarrier

The primary objective of validating simulants is to confirm mineralogical fidelity and simultaneously determine how their properties influence oxygen extraction behaviour and potential efficiency. Elemental mapping (SEM-EDX) and phase identification (XRD) establish baseline fidelity of the simulants. BET and TPR highlight textural and thermally responsive surface characteristics relevant to oxygen release and capture.

\subsection{Simulant Validation Outcomes}
%\FloatBarrier

\begin{figure}[t]
\centering
\includegraphics[width=\linewidth]{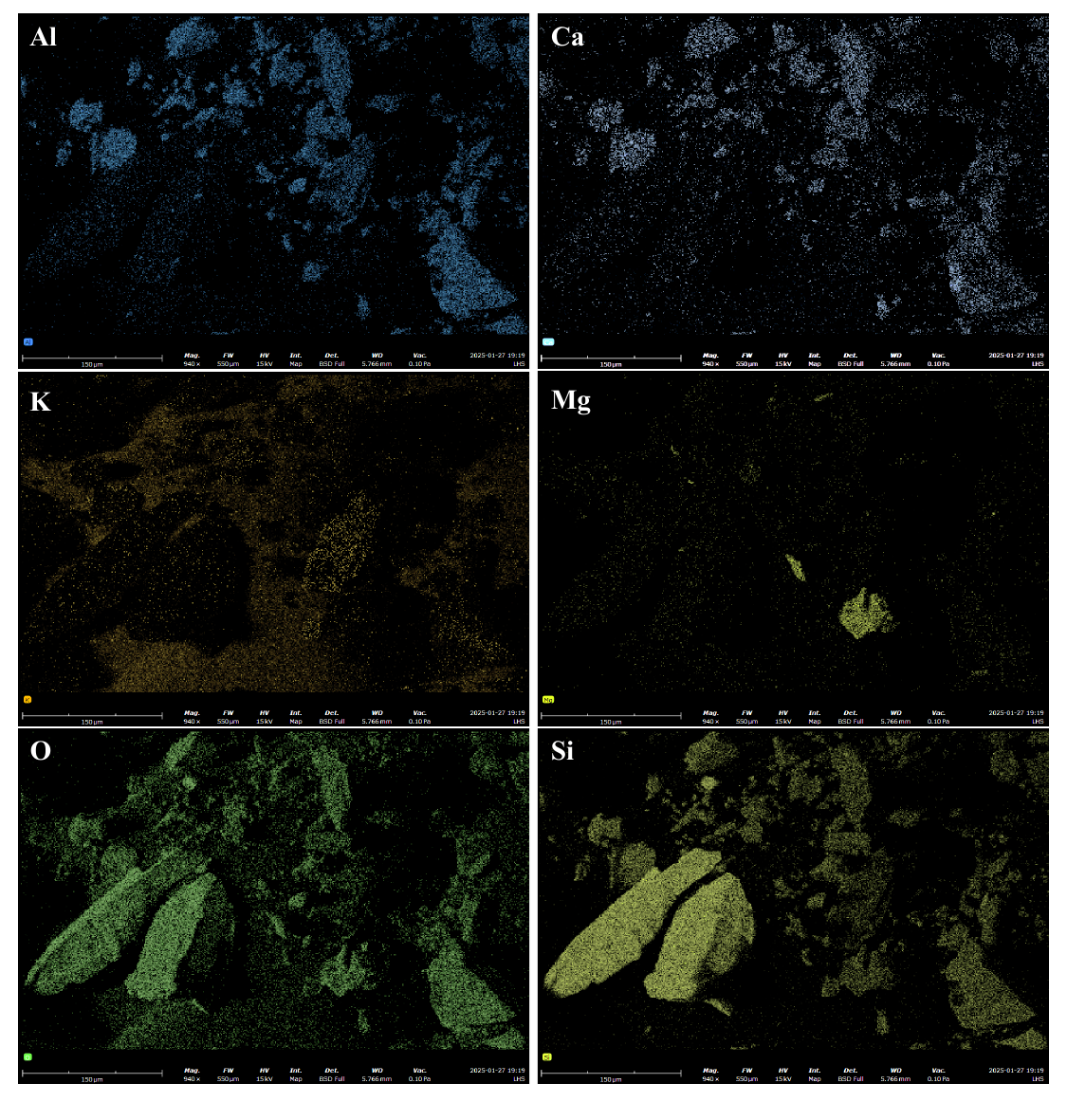}
\caption{EDX elemental map of LHS-2 simulant.}
\label{fig:lhs2_edx}
\end{figure}

Alongside an overall map of the elemental composition of LHS-2 and LSP-2, SEM-EDX can filter out respective maps of the individual elements present in each simulant.
As seen in Figure~\ref{fig:lhs2_edx}, the EDX Elemental Map shows large quantities of Si and O in LHS-2, and that these two elements are likely bonded as they are present in the same phases. Silicon and oxygen being dominant elements reflects the prevalence of anorthosite and basalt (SiO$_2$). This composition aligns with the mineralogical framework of lunar regolith, which is dominated by feldspathic and basaltic lithologies {\cite{mckay_properties_1990,zhang_size_2021}}. Anorthosites, forming the primary highland crust, consist mainly of Ca-rich plagioclase (An93-97) with minor mafic phases, originating through fractional crystallization and buoyant rise during Lunar Magma Ocean solidification \cite{wakita_lunar_1970,pernet-fisher_assessing_2017}. In contrast, mare basalts represent mantle-derived melts enriched in pyroxene and ilmenite, exhibiting higher FeO and TiO$_2$ contents, as documented in Apollo petrographic studies \cite{isaacson_lunar_2011} and recent Chang’e basalt analyses \cite{yin_petrogenesis_2025}. We confirmed the ilmenite presence in LHS-1 and LHS-2 through XRD phase identification and found it to be consistent with the reported simulant composition, despite its low overall abundance.

Thus preliminarily, SEM-EDX indicates that the elemental profile of LHS-2 provides a good match to lunar highlands regolith, supporting its suitability as a simulant for testing towards ISRU.

\begin{figure}[t]
\centering
\includegraphics[width=\linewidth]{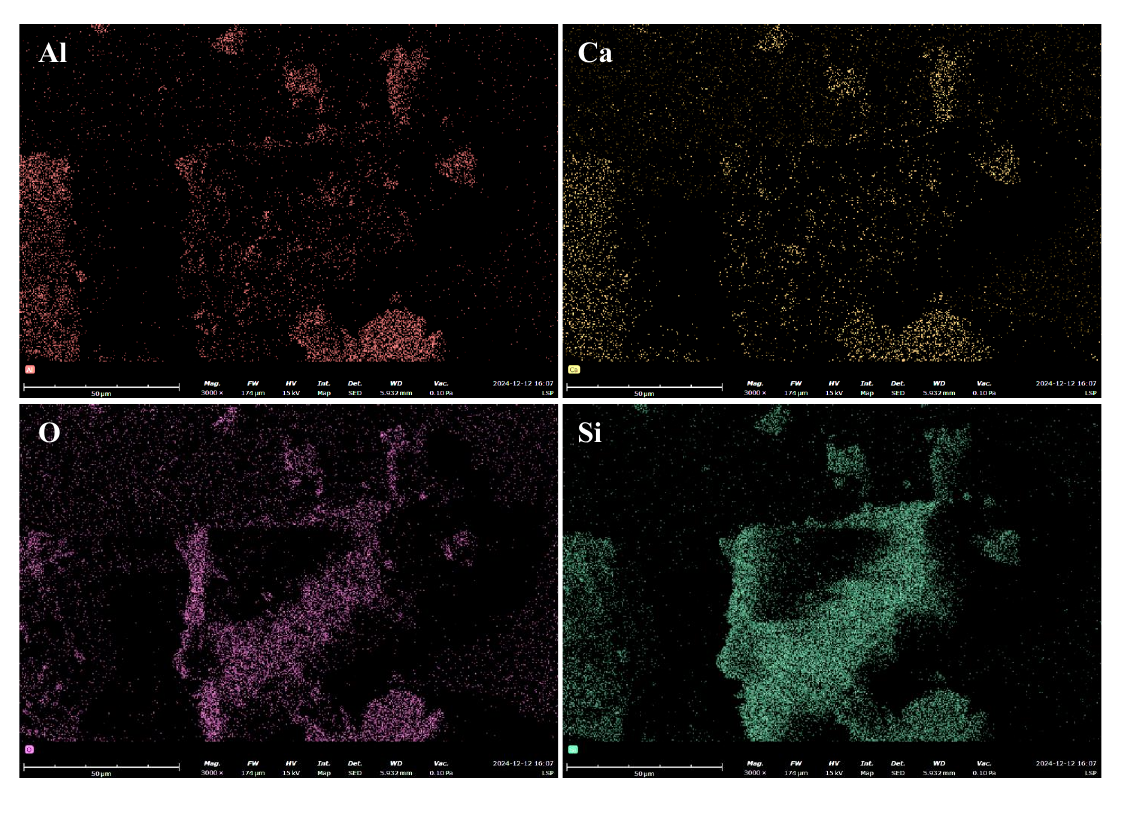}
\caption{EDX elemental map of LSP-2 simulant.}
\label{fig:lsp2_edx}
\end{figure}

Figure~\ref{fig:lsp2_edx} also presents the EDX Elemental Maps for LSP-2, revealing a consistent concentration of silicon (Si) and oxygen (O) within identical phases. These elemental distributions confirm the presence of basalt and anorthosite, consistent with the mineralogical mixed feldspathic-basaltic composition observed in the South Pole-Aitken basin \cite{zhang_provenance_2025}. Moreover, recent analyses from the Chang’e-6 mission indicate that the regolith in this region consists of approximately 93\% local mare basalt, with minor contributions from highland anorthositic and mantle-derived materials \cite{cui_sample_2024}.

\begin{table}[b]
\centering
\caption{Elemental concentrations of LHS-2 and LSP-2 simulants.}
\label{tab:edx_data}
\begin{tabular}{lcccc}
\toprule
Element &
\begin{tabular}[c]{@{}c@{}}LHS-2\\(at.\%)\end{tabular} &
\begin{tabular}[c]{@{}c@{}}LHS-2\\(wt\%)\end{tabular} &
\begin{tabular}[c]{@{}c@{}}LSP-2\\(at.\%)\end{tabular} &
\begin{tabular}[c]{@{}c@{}}LSP-2\\(wt\%)\end{tabular} \\
\midrule
O  & 74.418 & 61.486 & 74.749 & 61.924 \\
Si & 16.419 & 23.818 & 4.795  & 6.695  \\
Al & 5.458  & 7.601  & 17.836 & 25.942 \\
Ca & 2.529  & 5.236  & 2.621  & 5.439  \\
K  & 0.502  & 1.014  & --     & --     \\
Mg & 0.673  & 0.845  & --     & --     \\
\bottomrule
\end{tabular}
\end{table}

To verify the qualitative results of the imaging, Table~\ref{tab:edx_data} presents the EDX-derived elemental concentrations for LHS-2 and LSP-2. As expected, both materials exhibit high levels of oxygen and silicon, consistent with the silicate-rich nature of lunar regolith. Oxygen dominates in both samples ($\sim74$~at.\%), reflecting its role as a major component in oxides and silicate minerals \cite{robinot_review_2025}. This also aligns with the initially reported composition provided by the simulant manufacturer, as shown in Table~\ref{tab:reported_minerals}.

LHS-2 shows a notably high silicon content (16.4 at.\%, 23.8 wt\%), which aligns well with the mineralogy of lunar highland soils, where plagioclase feldspar and pyroxene are prevalent \cite{taylor_composition_1972}. This composition is again comparable to retrieved regolith samples. In contrast, LSP-2 displays ample aluminium levels (17.8 at.\%, 25.9 wt\%), suggesting presence of aluminosilicate phases common in lunar highland materials. Lower silicon concentrations in LSP-2 (4.8 at.\%, 6.7 wt\%) mimicking feldspathic lunar terrains.

Minor amounts of calcium, potassium, and magnesium are present in LHS-2, and the latter two are absent or below detection in LSP-2. These variations are within the expected range and reflect the diversity of regolith compositions observed across different lunar regions \cite{mckay_properties_1990}.

Overall, SEM-EDX both qualitatively and quantitatively suggests that LHS-2 and LSP-2 have well-aligned elemental profiles with lunar regolith, supporting the use of high-fidelity simulants for ISRU research. Additional details of the SEM–EDX characterization of the individual minerals can be found in the Appendix.

\begin{figure}[t]
\centering
\includegraphics[width=\linewidth]{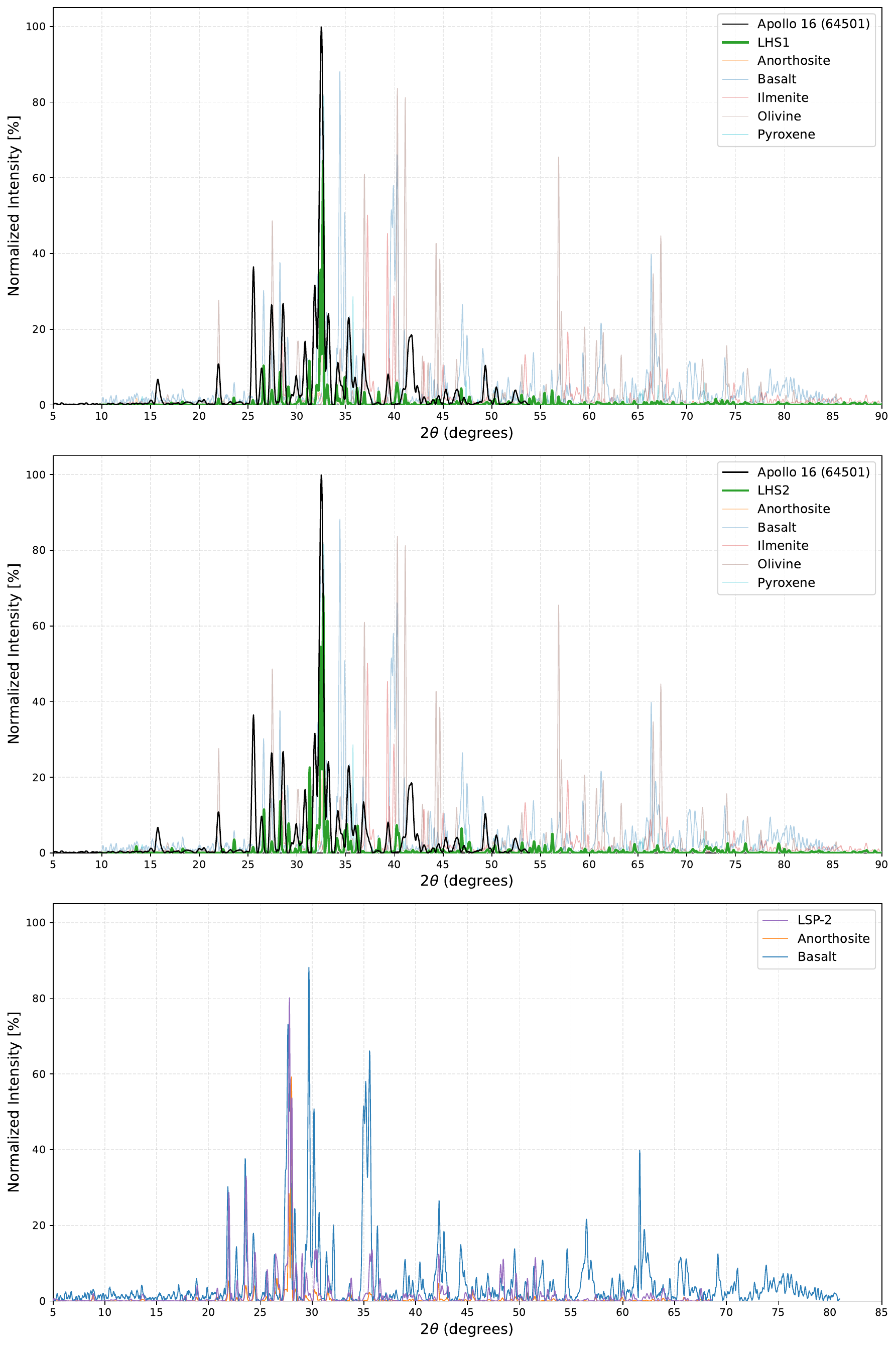}
\caption{XRD patterns of LHS-1 (top), LHS-2 (middle), LSP-2 (bottom), and their constituent minerals.}
\label{fig:xrd}
\end{figure}

As in Figure~\ref{fig:xrd}, XRD of the as-received powders enabled comparison of LHS-1 diffraction profiles against lunar-relevant minerals, including anorthosite, basalt, ilmenite, olivine, and pyroxene. LHS-1 (green) shows strongly-overlapping peaks with anorthosite (orange) and pyroxene (cyan), indicating significant presence of plagioclase and pyroxene phases which are indeed dominant in lunar highland regolith \cite{mckay1991lunar}. Sample 64501 (black) from the Apollo 16 mission shows substantial overlap, with minor deviations likely attributable to factors such as particle size distribution and crystallite effects \cite{li_preparation_2022}.

The middle panel illustrates the mineralogical profile of LHS-2. Its diffraction pattern (green) similarly aligns closely with anorthosite (orange) and pyroxene (cyan), confirming the presence of plagioclase and pyroxene phases consistent with lunar highland materials. Minor peak overlaps with ilmenite, olivine, and basalt suggest trace amounts of Fe-Ti oxides and mafic silicates, further supporting the simulant’s fidelity \cite{robinot_review_2025}. The resulting patterns for LHS-1 and LHS-2 exhibit strong correspondence in peak positions and relative intensities with Apollo 16, confirming their high mineralogical fidelity.

Finally, the bottom XRD pattern in Figure~\ref{fig:xrd} compares the diffraction profile of LSP-2 (purple) with those of anorthosite (orange) and basalt (blue), two key mineralogical components of lunar regolith. The strong peak alignment between LSP-2 and anorthosite suggests a dominant presence of plagioclase feldspar, characteristic of lunar highland materials. Additionally, the overlap with basalt indicates the inclusion of mafic phases, typical of mare regions \cite{salem_investigation_nodate}.

\subsection{Oxygen-Bearing Phases and Reduction Behavior}
%\FloatBarrier

\subsubsection{Hydrogen TPR Reduction Behavior}

Ilmenite (FeTiO$_3$) emerges as the most efficient oxygen source among lunar minerals, delivering an apparent yield of 1.10 wt\% oxygen under hydrogen reduction conditions. This result is aligned with work by NASA’s Applied Chemistry Laboratory \cite{muscatello_oxygen_2017} and more recently the ProSPA custom setup by Sargeant et al. \cite{sargeant_feasibility_2020} as well as Reiss et al. \cite{reiss_thermogravimetric_2019}, all of which demonstrated ilmenite-rich regolith yields up to 1-2 wt\% oxygen via hydrogen reduction at ~900 °C, following the reaction:

\begin{equation}
\mathrm{FeTiO_3 + H_2 \rightarrow Fe + TiO_2 + H_2O}
\end{equation}

This validates ilmenite as the most efficient known oxygen source and supports the use of high-fidelity simulants for research purposes. The reported weight percentages fall within the range documented in prior literature \cite{sargeant_hydrogen_2020}.

The corresponding TPR profiles for LHS, LSP, and ilmenite are illustrated in Figure 5. The observed reduction behavior in TPR is consistent with the presence of reducible Fe-Ti oxides such as ilmenite identified by XRD, even at low bulk concentrations.

\begin{figure}[htbp]
\centering
\includegraphics[width=\linewidth]{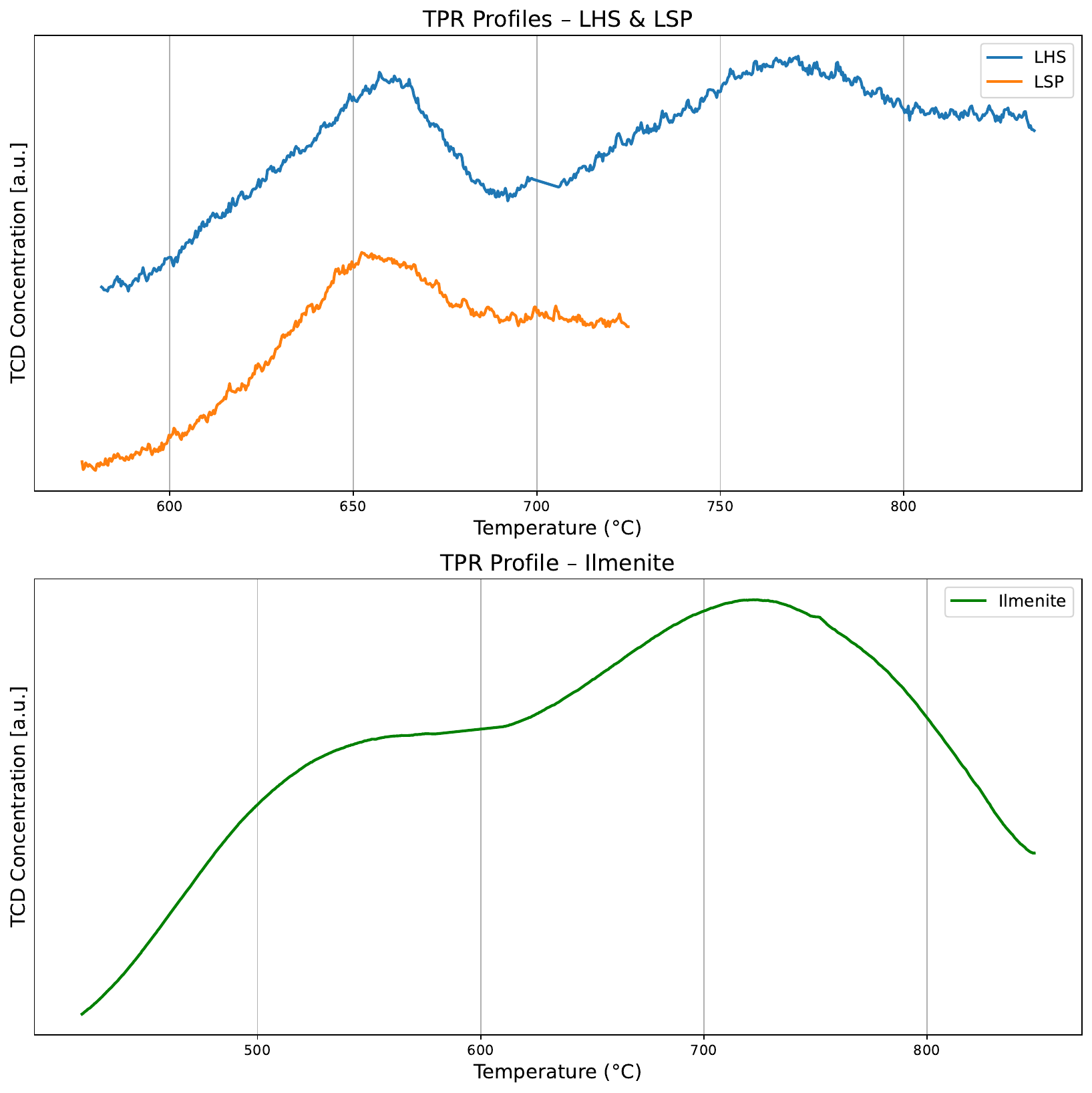}
\caption{TPR Profiles of LHS and LSP (top), Ilmenite (bottom).}
\label{fig:tpr_profiles}
\end{figure}

Among the mineralogical constituents of lunar regolith, Lunar Highlands Soil (LHS) emerges as a practical secondary source for in situ oxygen production due to its widespread distribution across the lunar surface. This is despite its low oxygen yield of approximately 0.02 wt\% inferred from TPR, the volumetric prevalence of LHS may compensate for its limited extraction efficiency. In addition, the Lunar South Pole (LSP) simulant exhibited an oxygen extraction of 0.01 wt\%. These values were obtained via TPR by baseline-correcting the TCD signal, integrating it over the reduction interval, and converting the area to oxygen quantity using a calibrated response normalized by sample mass.

Although TPR is widely used for catalyst characterization and oxygen quantification, the technique has inherent limitations \cite{dabbawala_toward_2023}. Baseline drift in thermal conductivity detectors can distort signal interpretation, while gas-composition cross-sensitivities may interfere with accurate hydrogen detection during TPR measurements \cite{munteanu_errors_2008}. Variations in heating rate strongly influence peak resolution and reduction profiles. Moreover, the reduction kinetics are highly influenced by the physicochemical characteristics of the material, particularly factors such as particle size, crystallite dimensions, and specific surface area \cite{baysal_study_2021}.

It is also important to note that the reduction pathway using hydrogen (H$_2$) produces water (H$_2$O) as an intermediate, rather than gaseous oxygen directly. While this introduces an additional processing step, water electrolysis, to liberate breathable O$_2$, it also presents a strategic advantage. Water is a dual-use resource: it can be split into hydrogen and oxygen for rocket fuel (LH$_2$/LOX), and used directly for life support, agriculture, and radiation shielding \cite{sargeant_hydrogen_2020,wang_review_2025}. Moreover, the hydrogen used in the reduction reaction can be recycled from the electrolysis step, enabling a closed-loop process that conserves hydrogen – a critical consideration given its scarcity on the Moon \cite{tewes_isru-derived_2020}.

Water is also easier to store, transport, and handle than gaseous oxygen, especially in cryogenic or pressurized systems. However, electrolysis can be power-demanding, as reduced gravity conditions have been shown to lower its efficiency and potentially limit its practicality within a mission’s available energy resource \cite{burke_modeling_2024}. The choice between this method and a direct oxygen extraction process (e.g., molten regolith electrolysis) will therefore depend on the specific mission architecture, available energy, and the desired product portfolio \cite{guerrero-gonzalez_system_2023,schreiner_molten_2015}.

\subsubsection{Textural Properties and Gas-Solid Accessibility}
%\FloatBarrier

\begin{table}[b]
\centering
\caption{Textural properties of lunar regolith simulant materials measured at 77~K.}
\label{tab:bet_results}
\small 
\setlength{\tabcolsep}{4pt} 
\begin{tabular}{lccc}
\toprule
Sample &
\begin{tabular}[c]{@{}c@{}}Surface area\\(m$^{2}$\,g$^{-1}$)\end{tabular} &
\begin{tabular}[c]{@{}c@{}}Pore volume\\(cm$^{3}$\,g$^{-1}$)\end{tabular} &
\begin{tabular}[c]{@{}c@{}}Average pore\\size (nm)\end{tabular} \\
\midrule
Anorthosite & 0.572 & 0.0195 & 136.53 \\
Basalt      & 0.599 & 0.0133 & 88.55  \\
Ilmenite    & 3.195 & 0.0168 & 21.05  \\
LHS-2       & 0.349 & 0.0151 & 26.83  \\
Olivine     & 1.056 & 0.0107 & 40.68  \\
Pyroxene    & 0.388 & 0.0131 & 135.52 \\
\bottomrule
\end{tabular}
\end{table}

Beyond chemical reducibility, physical and surface properties also influence gas-solid interactions relevant to oxygen extraction. Table~\ref{tab:bet_results} summarizes the textural properties of the lunar regolith simulants measured by N$_2$ adsorption-desoprtion isotherm at 77~K. Ilmenite exhibits the highest specific surface area (3.195~m$^2$/g) and mesopore-scale average pore size (21.05~nm), indicating comparatively high surface accessibility for gas-solid interactions. While BET-derived surface area and pore metrics do not directly predict oxide oxygen yield, they provide useful descriptors of surface accessibility and transport pathways that can influence gas-solid reaction kinetics, particularly during early reaction stages or under conditions where surface access is rate-limiting \cite{bui_determination_2022}.

Anorthosite and pyroxene exhibit relatively low surface areas (0.572 and 0.388 m²/g, respectively) with large average pore sizes ($\sim$ 135 nm), indicating a greater fraction of pore volume in larger voids rather than high internal surface area. Basalt and LHS-2 show similarly low surface areas (0.599 and 0.349 m²/g, respectively); notably, LHS-2 exhibits mesopore-scale average pore size (26.83 nm) despite its low BET area, consistent with limited accessible internal surface area in the bulk simulant due to aggregation, pore blocking, or packing effects in mixed-phase powders. Olivine exhibits intermediate textural properties (1.056 m²/g; 40.68 nm), consistent with moderate surface accessibility relative to the other silicate phases.

\subsubsection{Gas Adsorption Behavior and ISRU-Relevant Implications}

\begin{figure}[t]
\centering
\includegraphics[width=\linewidth]{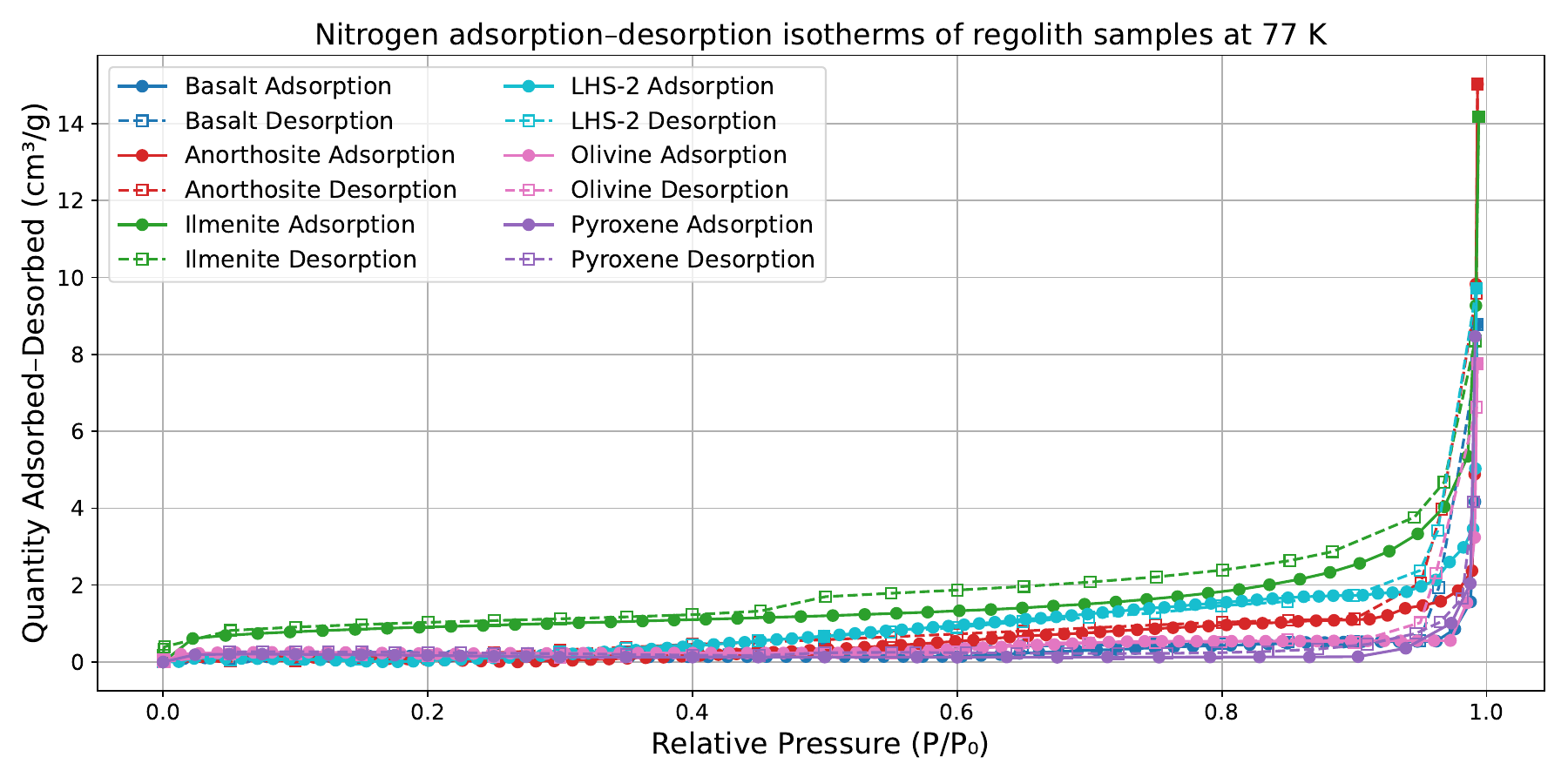}
\caption{Nitrogen adsorption-desorption isotherm of lunar regolith simulant materials at 77~K.}
\label{fig:isotherm}
\end{figure}

Nitrogen adsorption-desorption isotherms (Figure~\ref{fig:isotherm}) reveal distinct adsorption behaviors among the simulants, indicating that factors beyond specific surface area (i.e., surface chemistry, activation state, pore structure accessibility and distribution differences), play an important role in determining gas affinity {\cite{ma_advances_2025}}. Such behavior is relevant for ISRU concepts involving non-thermal plasma processing, volatile capture under vacuum conditions, and intermediate gas-handling steps following reduction or electrolysis, where gas-solid interactions can influence process performance.

\begin{table*}[htb]
\centering
\caption{Gas adsorption properties of lunar regolith simulant materials at 25~$^\circ$C.}
\label{tab:adsorption}
\setlength{\tabcolsep}{4.5pt} % default ~6pt; tighten to fit within margins
\begin{tabular}{lccccccc}
\toprule
Sample &
CO$_2$ & N$_2$ & H$_2$ & CH$_4$ &
CO$_2$/N$_2$ & CO$_2$/H$_2$ & CO$_2$/CH$_4$ \\
&
(mmol/g) & (mmol/g) & (mmol/g) & (mmol/g) & & & \\
\midrule
Anorthosite & 0.042 & 0.0019 & 0.0034 & --     & 22.11 & 12.35 & --   \\
Basalt      & 0.033 & 0.0018 & 0.0073 & --     & 18.33 & 4.52  & --   \\
Ilmenite    & 0.018 & 0.0083 & 0.0049 & --     & 2.17  & 3.67  & --   \\
LHS-2       & 0.044 & 0.0160 & 0.0063 & 0.0142 & 2.75  & 6.98  & 3.10 \\
Olivine     & 0.012 & 0.0029 & 0.0024 & --     & 4.14  & 5.00  & --   \\
Pyroxene    & 0.022 & 0.0063 & 0.0145 & 0.0068 & 3.49  & 1.52  & 3.24 \\
\bottomrule
\end{tabular}
\end{table*}

Collectively, these results indicate that while ilmenite remains the most reactive phase for chemical reduction processes, LHS-2 may offer advantages in adsorption- or capture-relevant contexts due to its mineralogical complexity and pore-structure characteristics under ISRU-relevant conditions \cite{petkov_comparison_2024}.

Selectivity values reported in Table~\ref{tab:adsorption} correspond to simple uptake ratios ($q_{\mathrm{CO_2}}/q_{\mathrm{gas}}$) derived from single-component adsorption at 1~bar and 25$^\circ$C, rather than thermodynamic or mixture-based selectivities.

LHS-2 exhibits the highest absolute CO$_2$ uptake (0.044~mmol/g), slightly exceeding anorthosite (0.042~mmol/g), despite having a comparatively low BET surface area. This indicates that CO$_2$ adsorption in LHS-2 is governed primarily by surface chemistry and microstructural heterogeneity rather than surface area alone. In contrast, anorthosite displays the highest CO$_2$/N$_2$ uptake ratio, which arises from very low N$_2$ adsorption rather than exceptionally high CO$_2$ capacity. Basalt similarly shows elevated CO$_2$/N$_2$ ratios driven by limited N$_2$ uptake. Ilmenite exhibits low CO$_2$ uptake combined with comparatively higher N$_2$ adsorption, resulting in low CO$_2$/N$_2$ selectivity, consistent with its limited affinity toward CO$_2$ under these conditions. Additional gas adsorption isotherm results at 25$^\circ$C and pressures up to 1~bar for CH$_4$, N$_2$, H$_2$, and CO$_2$ on the simulants are provided in the Appendix.

\begin{figure}[t]
\centering
\includegraphics[width=\linewidth]{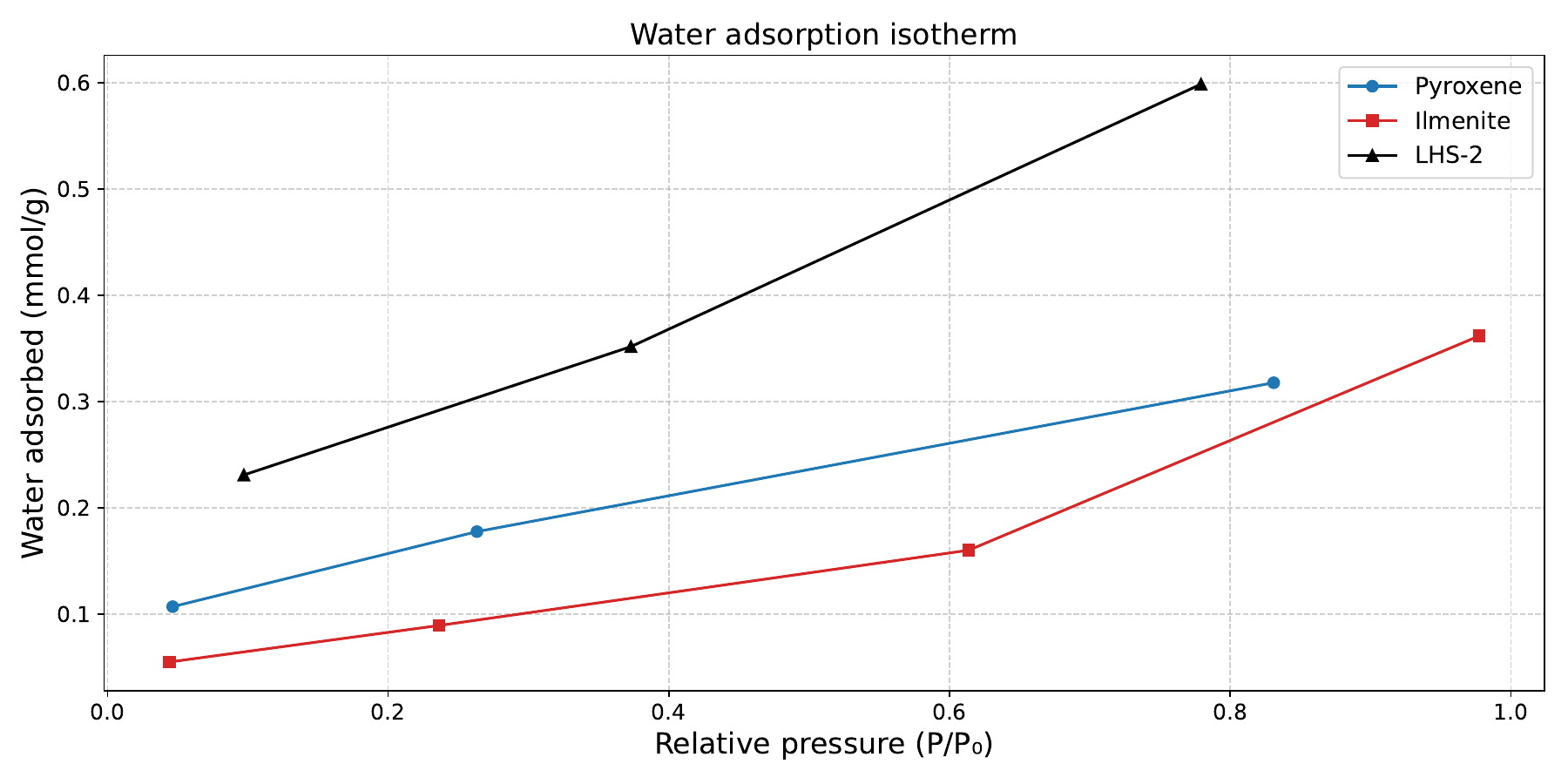}
\caption{Water adsorption isotherm of lunar regolith simulant materials at 25~$^\circ$C up to 100\% RH.}
\label{fig:water_isotherm}
\end{figure}

LHS-2 shows moderate CO$_2$ selectivity over N$_2$, H$_2$, and CH$_4$, reflecting high absolute adsorption capacity accompanied by relatively non-discriminatory uptake of multiple gases. This behavior is consistent with its mineralogical complexity and heterogeneous surface chemistry, which favor multi-gas adsorption rather than strong molecular selectivity.

Adsorption of H$_2$ and CH$_4$ is generally low across most samples; however, pyroxene exhibits comparatively high H$_2$ uptake and moderate CH$_4$ adsorption, suggesting that its pore structure or surface chemistry preferentially accommodates smaller gas molecules under these measurement conditions.

Figure \ref{fig:water_isotherm} shows the water adsorption isotherm of the materials at 25 °C up to 100\% relative humidity.

\section{Conclusion and Recommendations}
%\FloatBarrier

This study validates and characterizes high-fidelity lunar regolith simulants representative of the lunar highlands and south polar regions, with the objective of informing oxygen extraction strategies under realistic ISRU conditions. Combined XRD and SEM-EDX analyses confirm that the mineralogical and elemental profiles of the LHS-1, LHS-2, and LSP-2 simulants are broadly representative of returned Apollo and Chang’e samples, supporting their use as experimental analogs for lunar regolith in laboratory-scale ISRU studies.

Hydrogen temperature-programmed reduction confirms ilmenite as the most readily reducible oxygen-bearing phase, consistent with prior work. However, a key finding of this study is that bulk highland simulants, particularly LHS-2, exhibit favorable apparent oxygen release and gas-solid interaction behavior despite low ilmenite abundance. This behavior arises from the collective contribution of distributed Fe-bearing silicate and glassy phases, as well as surface and microstructural properties that influence reduction and adsorption processes. These results indicate that oxygen extraction behavior in realistic lunar regolith is governed by whole-regolith response rather than ilmenite content alone.

BET and gas adsorption analyses show that individual mineral phases have low surface areas, with ilmenite offering high accessibility but limited CO$_2$ affinity and silicate gas uptake not scaling with surface area. These results highlight the importance of mineralogical complexity and microstructural heterogeneity in governing gas–solid interactions relevant to ISRU processing. In this context, whole-regolith processing approaches may offer advantages in system simplicity and robustness compared to strategies relying on extensive mineral beneficiation, particularly for lunar highland and polar deployments where ilmenite is scarce \cite{shaw_mineral_2022}.

A key limitation of current simulants, as well as returned lunar samples, is their inability to fully replicate the space-weathered lunar surface, which is shaped by solar wind irradiation, micrometeoroid bombardment, and volatile retention \cite{l_vander_wal_lunar_2008,denevi_space_2023}. Addressing these effects will be essential for further improving the fidelity of ISRU testing, with ultimate validation requiring in situ experimentation on the lunar surface.

Based on these findings, we recommend several priorities for future research: (i) integrating bulk-regolith oxygen extraction behavior with reactor-scale and system-level analyses to assess process performance and operational trade-offs; (ii) developing next-generation simulants that incorporate space-weathering effects informed by data from lunar polar missions; and (iii) employing high-fidelity simulants to evaluate emerging ISRU approaches, such as non-thermal plasma-based methods \cite{alhemeiri_advancing_2024}, alongside continued optimization of established processes including ilmenite reduction \cite{zou_development_2024}. In addition, further investigation of the mineralogical and textural properties of lunar regolith may clarify its potential catalytic or kinetic role in facilitating gas-solid reactions relevant to future ISRU applications \cite{zhong_situ_2023}.

In summary, while no simulant can fully reproduce the lunar environment, this study demonstrates the value of high-fidelity regolith analogs for bridging laboratory experimentation and lunar ISRU system design. The insights gained, particularly regarding the role of whole-regolith behavior, provide a foundation for developing robust and adaptable oxygen extraction strategies needed to support sustained human presence beyond Earth.

\section*{Acknowledgements}
A.A.D., L.K., and M.G. acknowledge Khalifa University of Science and Technology for funding through projects \#8474000740-RIG-2024-045 and \#8474000486-ESIG-2023-015. The authors acknowledge Khalifa University for providing facilities and institutional support, and the CeCaS Fund (RC2-2018-024) for access to facilities and infrastructure.

\appendix
\nocite{Monshi2012}

\bibliographystyle{elsarticle-harv}
\bibliography{references}

\begin{thebibliography}{50}
\expandafter\ifx\csname natexlab\endcsname\relax\def\natexlab#1{#1}\fi
\providecommand{\url}[1]{\texttt{#1}}
\providecommand{\href}[2]{#2}
\providecommand{\path}[1]{#1}
\providecommand{\DOIprefix}{doi:}
\providecommand{\ArXivprefix}{arXiv:}
\providecommand{\URLprefix}{URL: }
\providecommand{\Pubmedprefix}{pmid:}
\providecommand{\doi}[1]{\href{http://dx.doi.org/#1}{\path{#1}}}
\providecommand{\Pubmed}[1]{\href{pmid:#1}{\path{#1}}}
\providecommand{\bibinfo}[2]{#2}
\ifx\xfnm\relax \def\xfnm[#1]{\unskip,\space#1}\fi
%Type = Article
\bibitem[{Alhemeiri et~al.(2024)Alhemeiri, Kosca, Gacesa and Polychronopoulou}]{alhemeiri_advancing_2024}
\bibinfo{author}{Alhemeiri, N.}, \bibinfo{author}{Kosca, L.}, \bibinfo{author}{Gacesa, M.}, \bibinfo{author}{Polychronopoulou, K.}, \bibinfo{year}{2024}.
\newblock \bibinfo{title}{Advancing in-situ resource utilization for earth and space applications through plasma {CO2} catalysis}.
\newblock \bibinfo{journal}{Journal of CO2 Utilization} \bibinfo{volume}{85}, \bibinfo{pages}{102887}.
\newblock \DOIprefix\doi{10.1016/j.jcou.2024.102887}.
%Type = Article
\bibitem[{Baysal et~al.(2021)Baysal, Kirchner, Mehne and Kureti}]{baysal_study_2021}
\bibinfo{author}{Baysal, Z.}, \bibinfo{author}{Kirchner, J.}, \bibinfo{author}{Mehne, M.}, \bibinfo{author}{Kureti, S.}, \bibinfo{year}{2021}.
\newblock \bibinfo{title}{Study on the reduction of ilmenite-type {FeTiO3} by {H2}}.
\newblock \bibinfo{journal}{International Journal of Hydrogen Energy} \bibinfo{volume}{46}, \bibinfo{pages}{4447--4459}.
\newblock \DOIprefix\doi{10.1016/j.ijhydene.2020.10.266}.
%Type = Article
\bibitem[{Bui et~al.(2022)Bui, Nagapudi and Chakravarty}]{bui_determination_2022}
\bibinfo{author}{Bui, M.}, \bibinfo{author}{Nagapudi, K.}, \bibinfo{author}{Chakravarty, P.}, \bibinfo{year}{2022}.
\newblock \bibinfo{title}{Determination of {BET} {Specific} {Surface} {Area} of {Hydrate}-{Anhydrate} {Systems} {Susceptible} to {Phase} {Transformation} {Using} {Inverse} {Gas} {Chromatography}}.
\newblock \bibinfo{journal}{AAPS PharmSciTech} \bibinfo{volume}{23}, \bibinfo{pages}{237}.
\newblock \DOIprefix\doi{10.1208/s12249-022-02395-6}.
%Type = Article
\bibitem[{Burke et~al.(2024)Burke, Nord, Hibbitts and Berdis}]{burke_modeling_2024}
\bibinfo{author}{Burke, P.}, \bibinfo{author}{Nord, M.}, \bibinfo{author}{Hibbitts, C.}, \bibinfo{author}{Berdis, J.}, \bibinfo{year}{2024}.
\newblock \bibinfo{title}{Modeling electrolysis in reduced gravity: producing oxygen from in-situ resources at the moon and beyond}.
\newblock \bibinfo{journal}{Frontiers in Space Technologies} \bibinfo{volume}{5}.
\newblock \DOIprefix\doi{10.3389/frspt.2024.1304579}.
%Type = Article
\bibitem[{Cui et~al.(2024)Cui, Yang, Zhang, Wang, Xian, Chen, Xiao, Qian, Head, Neal, Xiao, Luo, Chen, He, Cao, Zhou, Huang, Chen, Wei, Wang, Yang, Li, Yang, Lin, Zhu, Zhang and Xu}]{cui_sample_2024}
\bibinfo{author}{Cui, Z.}, \bibinfo{author}{Yang, Q.}, \bibinfo{author}{Zhang, Y.Q.}, \bibinfo{author}{Wang, C.}, \bibinfo{author}{Xian, H.}, \bibinfo{author}{Chen, Z.}, \bibinfo{author}{Xiao, Z.}, \bibinfo{author}{Qian, Y.}, \bibinfo{author}{Head, J.W.}, \bibinfo{author}{Neal, C.R.}, \bibinfo{author}{Xiao, L.}, \bibinfo{author}{Luo, F.}, \bibinfo{author}{Chen, J.}, \bibinfo{author}{He, P.}, \bibinfo{author}{Cao, Y.}, \bibinfo{author}{Zhou, Q.}, \bibinfo{author}{Huang, F.}, \bibinfo{author}{Chen, L.}, \bibinfo{author}{Wei, B.}, \bibinfo{author}{Wang, J.}, \bibinfo{author}{Yang, Y.N.}, \bibinfo{author}{Li, S.}, \bibinfo{author}{Yang, Y.}, \bibinfo{author}{Lin, X.}, \bibinfo{author}{Zhu, J.}, \bibinfo{author}{Zhang, L.}, \bibinfo{author}{Xu, Y.G.}, \bibinfo{year}{2024}.
\newblock \bibinfo{title}{A sample of the {Moon}’s far side retrieved by {Chang}’e-6 contains 2.83-billion-year-old basalt}.
\newblock \bibinfo{journal}{Science} \bibinfo{volume}{386}, \bibinfo{pages}{1395--1399}.
\newblock \DOIprefix\doi{10.1126/science.adt1093}. \bibinfo{note}{publisher: American Association for the Advancement of Science}.
%Type = Misc
\bibitem[{D.~Green and Kleinhenz(2019)}]{d_green_-situ_2019}
\bibinfo{author}{D.~Green, R.}, \bibinfo{author}{Kleinhenz, J.}, \bibinfo{year}{2019}.
\newblock \bibinfo{title}{In-{Situ} {Resource} {Utilization} ({ISRU}) {Living} off the {Land}}.
%Type = Article
\bibitem[{Dabbawala et~al.(2023)Dabbawala, Elmutasim, Baker, Siakavelas, Anjum, Charisiou, Hinder, Munro, Gacesa, Goula and Polychronopoulou}]{dabbawala_toward_2023}
\bibinfo{author}{Dabbawala, A.A.}, \bibinfo{author}{Elmutasim, O.}, \bibinfo{author}{Baker, M.A.}, \bibinfo{author}{Siakavelas, G.}, \bibinfo{author}{Anjum, D.H.}, \bibinfo{author}{Charisiou, N.D.}, \bibinfo{author}{Hinder, S.J.}, \bibinfo{author}{Munro, C.J.}, \bibinfo{author}{Gacesa, M.}, \bibinfo{author}{Goula, M.A.}, \bibinfo{author}{Polychronopoulou, K.}, \bibinfo{year}{2023}.
\newblock \bibinfo{title}{Toward maximizing the selectivity of diesel-like hydrocarbons from oleic acid hydrodeoxygenation using {Ni}/{Co}-{Al2O3} embedded mesoporous silica nanocomposite catalysts: {An} experimental and {DFT} approach}.
\newblock \bibinfo{journal}{Applied Surface Science} \bibinfo{volume}{640}, \bibinfo{pages}{158294}.
\newblock \DOIprefix\doi{10.1016/j.apsusc.2023.158294}.
%Type = Article
\bibitem[{Denevi et~al.(2023)Denevi, Noble, Christoffersen, Thompson, Glotch, Blewett, Garrick-Bethell, Gillis-Davis, Greenhagen, Hendrix, Hurley, Keller, Kramer and Trang}]{denevi_space_2023}
\bibinfo{author}{Denevi, B.W.}, \bibinfo{author}{Noble, S.K.}, \bibinfo{author}{Christoffersen, R.}, \bibinfo{author}{Thompson, M.S.}, \bibinfo{author}{Glotch, T.D.}, \bibinfo{author}{Blewett, D.T.}, \bibinfo{author}{Garrick-Bethell, I.}, \bibinfo{author}{Gillis-Davis, J.J.}, \bibinfo{author}{Greenhagen, B.T.}, \bibinfo{author}{Hendrix, A.R.}, \bibinfo{author}{Hurley, D.M.}, \bibinfo{author}{Keller, L.P.}, \bibinfo{author}{Kramer, G.Y.}, \bibinfo{author}{Trang, D.}, \bibinfo{year}{2023}.
\newblock \bibinfo{title}{Space {Weathering} {At} {The} {Moon}}.
\newblock \bibinfo{journal}{Reviews in Mineralogy and Geochemistry} \bibinfo{volume}{89}, \bibinfo{pages}{611--650}.
\newblock \URLprefix \url{https://doi.org/10.2138/rmg.2023.89.14}, \DOIprefix\doi{10.2138/rmg.2023.89.14}. \bibinfo{note}{\_eprint: https://pubs.geoscienceworld.org/msa/rimg/article-pdf/89/1/611/6072958/rmg.2023.89.14.pdf}.
%Type = Article
\bibitem[{Easter et~al.(2024)Easter, Long-Fox, Britt and Brisset}]{easter_effect_2024}
\bibinfo{author}{Easter, P.}, \bibinfo{author}{Long-Fox, J.}, \bibinfo{author}{Britt, D.}, \bibinfo{author}{Brisset, J.}, \bibinfo{year}{2024}.
\newblock \bibinfo{title}{The effect of particle size distribution on lunar regolith simulant angle of repose}.
\newblock \bibinfo{journal}{Advances in Space Research} .
%Type = Article
\bibitem[{Guerrero-Gonzalez and Zabel(2023)}]{guerrero-gonzalez_system_2023}
\bibinfo{author}{Guerrero-Gonzalez, F.J.}, \bibinfo{author}{Zabel, P.}, \bibinfo{year}{2023}.
\newblock \bibinfo{title}{System analysis of an {ISRU} production plant: {Extraction} of metals and oxygen from lunar regolith}.
\newblock \bibinfo{journal}{Acta Astronautica} \bibinfo{volume}{203}, \bibinfo{pages}{187--201}.
\newblock \DOIprefix\doi{10.1016/j.actaastro.2022.11.050}.
%Type = Article
\bibitem[{Hulbert and Kriven(2023)}]{hulbert_specimen-displacement_2023}
\bibinfo{author}{Hulbert, B.S.}, \bibinfo{author}{Kriven, W.M.}, \bibinfo{year}{2023}.
\newblock \bibinfo{title}{Specimen-displacement correction for powder {X}-ray diffraction in {Debye}–{Scherrer} geometry with a flat area detector}.
\newblock \bibinfo{journal}{Journal of Applied Crystallography} \bibinfo{volume}{56}, \bibinfo{pages}{160--166}.
\newblock \DOIprefix\doi{10.1107/S1600576722011360}. \bibinfo{note}{publisher: International Union of Crystallography}.
%Type = Article
\bibitem[{Isaacson et~al.(2011)Isaacson, Sarbadhikari, Pieters, Klima, Hiroi, Liu and Taylor}]{isaacson_lunar_2011}
\bibinfo{author}{Isaacson, P.J.}, \bibinfo{author}{Sarbadhikari, A.B.}, \bibinfo{author}{Pieters, C.M.}, \bibinfo{author}{Klima, R.L.}, \bibinfo{author}{Hiroi, T.}, \bibinfo{author}{Liu, Y.}, \bibinfo{author}{Taylor, L.A.}, \bibinfo{year}{2011}.
\newblock \bibinfo{title}{The lunar rock and mineral characterization consortium: {Deconstruction} and integrated mineralogical, petrologic, and spectroscopic analyses of mare basalts}.
\newblock \bibinfo{journal}{Meteoritics \& Planetary Science} \bibinfo{volume}{46}, \bibinfo{pages}{228--251}.
\newblock \DOIprefix\doi{10.1111/j.1945-5100.2010.01148.x}.
%Type = Article
\bibitem[{Isachenkov et~al.(2022)Isachenkov, Chugunov, Landsman, Akhatov, Metke, Tikhonov and Shishkovsky}]{isachenkov_characterization_2022}
\bibinfo{author}{Isachenkov, M.}, \bibinfo{author}{Chugunov, S.}, \bibinfo{author}{Landsman, Z.}, \bibinfo{author}{Akhatov, I.}, \bibinfo{author}{Metke, A.}, \bibinfo{author}{Tikhonov, A.}, \bibinfo{author}{Shishkovsky, I.}, \bibinfo{year}{2022}.
\newblock \bibinfo{title}{Characterization of novel lunar highland and mare simulants for {ISRU} research applications}.
\newblock \bibinfo{journal}{Icarus} \bibinfo{volume}{376}, \bibinfo{pages}{114873}.
\newblock \DOIprefix\doi{10.1016/j.icarus.2021.114873}.
%Type = Techreport
\bibitem[{L.~Vander~Wal(2008)}]{l_vander_wal_lunar_2008}
\bibinfo{author}{L.~Vander~Wal, R.}, \bibinfo{year}{2008}.
\newblock \bibinfo{title}{Lunar {Dust} {Chemical}, {Electrical}, and {Mechanical} {Reactivity}: {Simulation} and {Characterization}}.
\newblock \bibinfo{type}{Technical Report} \bibinfo{number}{NASA/CR—2008-215431}. The National Center for Space Exploration Research, Glenn Research Center, Cleveland, Ohio.
%Type = Article
\bibitem[{Li et~al.(2022)Li, Zhou, Yan, Chen, Chen, Cai and Mo}]{li_preparation_2022}
\bibinfo{author}{Li, R.}, \bibinfo{author}{Zhou, G.}, \bibinfo{author}{Yan, K.}, \bibinfo{author}{Chen, J.}, \bibinfo{author}{Chen, D.}, \bibinfo{author}{Cai, S.}, \bibinfo{author}{Mo, P.Q.}, \bibinfo{year}{2022}.
\newblock \bibinfo{title}{Preparation and characterization of a specialized lunar regolith simulant for use in lunar low gravity simulation}.
\newblock \bibinfo{journal}{International Journal of Mining Science and Technology} \bibinfo{volume}{32}, \bibinfo{pages}{1--15}.
\newblock \DOIprefix\doi{10.1016/j.ijmst.2021.09.003}.
%Type = Article
\bibitem[{Long-Fox et~al.(2023)Long-Fox, Landsman, Easter, Millwater and Britt}]{long-fox_geomechanical_2023}
\bibinfo{author}{Long-Fox, J.M.}, \bibinfo{author}{Landsman, Z.A.}, \bibinfo{author}{Easter, P.B.}, \bibinfo{author}{Millwater, C.A.}, \bibinfo{author}{Britt, D.T.}, \bibinfo{year}{2023}.
\newblock \bibinfo{title}{Geomechanical properties of lunar regolith simulants {LHS}-1 and {LMS}-1}.
\newblock \bibinfo{journal}{Advances in Space Research} \bibinfo{volume}{71}, \bibinfo{pages}{5400--5412}.
\newblock \DOIprefix\doi{10.1016/j.asr.2023.02.034}.
%Type = Article
\bibitem[{Ma et~al.(2025)Ma, Fu, Tong, Umar, Hung and Wang}]{ma_advances_2025}
\bibinfo{author}{Ma, H.}, \bibinfo{author}{Fu, H.}, \bibinfo{author}{Tong, Y.}, \bibinfo{author}{Umar, A.}, \bibinfo{author}{Hung, Y.M.}, \bibinfo{author}{Wang, X.}, \bibinfo{year}{2025}.
\newblock \bibinfo{title}{Advances in {CO2} capture and separation materials: {Emerging} trends, challenges, and prospects for sustainable applications}.
\newblock \bibinfo{journal}{Carbon Capture Science \& Technology} \bibinfo{volume}{15}, \bibinfo{pages}{100441}.
\newblock \DOIprefix\doi{https://doi.org/10.1016/j.ccst.2025.100441}.
%Type = Incollection
\bibitem[{Mckay and Ming(1990)}]{mckay_properties_1990}
\bibinfo{author}{Mckay, D.}, \bibinfo{author}{Ming, D.}, \bibinfo{year}{1990}.
\newblock \bibinfo{title}{Properties of {Lunar} {Regolith}}, in: \bibinfo{booktitle}{Developments in {Soil} {Science}}. \bibinfo{publisher}{Elsevier}. volume~\bibinfo{volume}{19}, pp. \bibinfo{pages}{449--462}.
\newblock \DOIprefix\doi{10.1016/S0166-2481(08)70360-X}. \bibinfo{note}{iSSN: 0166-2481}.
%Type = Article
\bibitem[{McKay et~al.(1991)McKay, Heiken, Basu, Blanford, Simon, Reedy, French and Papike}]{mckay1991lunar}
\bibinfo{author}{McKay, D.S.}, \bibinfo{author}{Heiken, G.}, \bibinfo{author}{Basu, A.}, \bibinfo{author}{Blanford, G.}, \bibinfo{author}{Simon, S.}, \bibinfo{author}{Reedy, R.}, \bibinfo{author}{French, B.M.}, \bibinfo{author}{Papike, J.}, \bibinfo{year}{1991}.
\newblock \bibinfo{title}{The lunar regolith}.
\newblock \bibinfo{journal}{Lunar sourcebook} \bibinfo{volume}{567}, \bibinfo{pages}{285--356}.
%Type = Article
\bibitem[{Monshi et~al.(2012)Monshi, Foroughi and Monshi}]{Monshi2012}
\bibinfo{author}{Monshi, A.}, \bibinfo{author}{Foroughi, M.R.}, \bibinfo{author}{Monshi, M.R.}, \bibinfo{year}{2012}.
\newblock \bibinfo{title}{Modified scherrer equation to estimate more accurately nano-crystallite size using xrd}.
\newblock \bibinfo{journal}{World Journal of Nano Science and Engineering} \bibinfo{volume}{2}, \bibinfo{pages}{154--160}.
\newblock \DOIprefix\doi{10.4236/wjnse.2012.23020}.
%Type = Article
\bibitem[{Munteanu et~al.(2008)Munteanu, Miclea and Segal}]{munteanu_errors_2008}
\bibinfo{author}{Munteanu, G.}, \bibinfo{author}{Miclea, C.}, \bibinfo{author}{Segal, E.}, \bibinfo{year}{2008}.
\newblock \bibinfo{title}{Errors in evaluation of the kinetic parameters in temperature programmed reduction}.
\newblock \bibinfo{journal}{Journal of Thermal Analysis and Calorimetry} \bibinfo{volume}{94}, \bibinfo{pages}{317--321}.
\newblock \DOIprefix\doi{10.1007/s10973-008-8998-y}.
%Type = Inproceedings
\bibitem[{Muscatello(2017)}]{muscatello_oxygen_2017}
\bibinfo{author}{Muscatello, T.}, \bibinfo{year}{2017}.
\newblock \bibinfo{title}{Oxygen {Extraction} from {Minerals}}, in: \bibinfo{booktitle}{{NASA} {Applied} {Chemistry} {Laboratory}}.
%Type = Article
\bibitem[{Pernet-Fisher et~al.(2017)Pernet-Fisher, Joy, Martin and Donaldson~Hanna}]{pernet-fisher_assessing_2017}
\bibinfo{author}{Pernet-Fisher, J.F.}, \bibinfo{author}{Joy, K.H.}, \bibinfo{author}{Martin, D.J.P.}, \bibinfo{author}{Donaldson~Hanna, K.L.}, \bibinfo{year}{2017}.
\newblock \bibinfo{title}{Assessing the shock state of the lunar highlands: {Implications} for the petrogenesis and chronology of crustal anorthosites}.
\newblock \bibinfo{journal}{Scientific Reports} \bibinfo{volume}{7}, \bibinfo{pages}{5888}.
\newblock \DOIprefix\doi{10.1038/s41598-017-06134-x}. \bibinfo{note}{publisher: Nature Publishing Group}.
%Type = Article
\bibitem[{Petkov et~al.(2024)Petkov, Wilkerson, Voecks, Rickman, Edmunson and Effinger}]{petkov_comparison_2024}
\bibinfo{author}{Petkov, M.P.}, \bibinfo{author}{Wilkerson, R.P.}, \bibinfo{author}{Voecks, G.E.}, \bibinfo{author}{Rickman, D.L.}, \bibinfo{author}{Edmunson, J.E.}, \bibinfo{author}{Effinger, M.R.}, \bibinfo{year}{2024}.
\newblock \bibinfo{title}{Comparison of volatiles evolving from selected highland and mare lunar regolith simulants during vacuum sintering}.
\newblock \bibinfo{journal}{Planetary and Space Science} \bibinfo{volume}{252}, \bibinfo{pages}{105982}.
\newblock \DOIprefix\doi{10.1016/j.pss.2024.105982}.
%Type = Incollection
\bibitem[{Ramachandran et~al.(2006)Ramachandran, Willenberg and Volz}]{ramachandran_hydrogen_2006}
\bibinfo{author}{Ramachandran, N.}, \bibinfo{author}{Willenberg, H.}, \bibinfo{author}{Volz, M.}, \bibinfo{year}{2006}.
\newblock \bibinfo{title}{Hydrogen {Reduction} of {Ilmenite} from {Lunar} {Regolith}}, in: \bibinfo{booktitle}{Space 2006}. \bibinfo{publisher}{American Institute of Aeronautics and Astronautics}. {AIAA} {SPACE} {Forum}.
\newblock \DOIprefix\doi{10.2514/6.2006-7482}.
%Type = Article
\bibitem[{Reiss et~al.(2019)Reiss, Kerscher and Grill}]{reiss_thermogravimetric_2019}
\bibinfo{author}{Reiss, P.}, \bibinfo{author}{Kerscher, F.}, \bibinfo{author}{Grill, L.}, \bibinfo{year}{2019}.
\newblock \bibinfo{title}{Thermogravimetric {Analysis} of the {Reduction} of {Ilmenite} and {NU}-{LHT}-{2M} with {Hydrogen} and {Methane}}.
\newblock \bibinfo{journal}{Developing a New Space Economy} .
%Type = Article
\bibitem[{Robinot et~al.(2025)Robinot, Rodat, Abanades, Paillet and Cowley}]{robinot_review_2025}
\bibinfo{author}{Robinot, J.}, \bibinfo{author}{Rodat, S.}, \bibinfo{author}{Abanades, S.}, \bibinfo{author}{Paillet, A.}, \bibinfo{author}{Cowley, A.}, \bibinfo{year}{2025}.
\newblock \bibinfo{title}{Review of in-situ oxygen extraction from lunar regolith with focus on solar thermal and laser vacuum pyrolysis}.
\newblock \bibinfo{journal}{Acta Astronautica} \bibinfo{volume}{234}, \bibinfo{pages}{242--259}.
\newblock \DOIprefix\doi{10.1016/j.actaastro.2025.05.008}.
%Type = Article
\bibitem[{Salem et~al.()Salem, Sharma, Kumaresan, Karthi., Howari, Nazzal and Xavier}]{salem_investigation_nodate}
\bibinfo{author}{Salem, I.B.}, \bibinfo{author}{Sharma, M.}, \bibinfo{author}{Kumaresan, P.}, \bibinfo{author}{Karthi., A.}, \bibinfo{author}{Howari, F.M.}, \bibinfo{author}{Nazzal, Y.}, \bibinfo{author}{Xavier, C.M.}, .
\newblock \bibinfo{title}{An {Investigation} on the {Morphological} and {Mineralogical} {Characteristics} of {Posidonius} {Floor} {Fractured} {Lunar} {Impact} {Crater} {Using} {Lunar} {Remote} {Sensing} {Data}}.
\newblock \bibinfo{journal}{MDPI} \DOIprefix\doi{https://doi.org/10.3390/rs14040814}.
%Type = Article
\bibitem[{Sargeant et~al.(2020a)Sargeant, Abernethy, Anand, Barber, Landsberg, Sheridan, Wright and Morse}]{sargeant_feasibility_2020}
\bibinfo{author}{Sargeant, H.M.}, \bibinfo{author}{Abernethy, F.A.J.}, \bibinfo{author}{Anand, M.}, \bibinfo{author}{Barber, S.J.}, \bibinfo{author}{Landsberg, P.}, \bibinfo{author}{Sheridan, S.}, \bibinfo{author}{Wright, I.}, \bibinfo{author}{Morse, A.}, \bibinfo{year}{2020}a.
\newblock \bibinfo{title}{Feasibility studies for hydrogen reduction of ilmenite in a static system for use as an {ISRU} demonstration on the lunar surface}.
\newblock \bibinfo{journal}{Planetary and Space Science} \bibinfo{volume}{180}, \bibinfo{pages}{104759}.
\newblock \DOIprefix\doi{10.1016/j.pss.2019.104759}.
%Type = Article
\bibitem[{Sargeant et~al.(2020b)Sargeant, Abernethy, Barber, Wright, Anand, Sheridan and Morse}]{sargeant_hydrogen_2020}
\bibinfo{author}{Sargeant, H.M.}, \bibinfo{author}{Abernethy, F.A.J.}, \bibinfo{author}{Barber, S.J.}, \bibinfo{author}{Wright, I.P.}, \bibinfo{author}{Anand, M.}, \bibinfo{author}{Sheridan, S.}, \bibinfo{author}{Morse, A.}, \bibinfo{year}{2020}b.
\newblock \bibinfo{title}{Hydrogen reduction of ilmenite: {Towards} an in situ resource utilization demonstration on the surface of the {Moon}}.
\newblock \bibinfo{journal}{Planetary and Space Science} \bibinfo{volume}{180}, \bibinfo{pages}{104751}.
\newblock \DOIprefix\doi{10.1016/j.pss.2019.104751}.
%Type = Phdthesis
\bibitem[{Schreiner(2015)}]{schreiner_molten_2015}
\bibinfo{author}{Schreiner, S.S.}, \bibinfo{year}{2015}.
\newblock \bibinfo{title}{Molten {Regolith} {Electrolysis} reactor modeling and optimization of in-situ resource utilization systems}.
\newblock \bibinfo{type}{Thesis}. Massachusetts Institute of Technology.
\newblock \bibinfo{note}{Accepted: 2015-09-17T17:44:44Z}.
%Type = Article
\bibitem[{Shaw et~al.(2022)Shaw, Humbert, Brooks, Rhamdhani, Duffy and Pownceby}]{shaw_mineral_2022}
\bibinfo{author}{Shaw, M.}, \bibinfo{author}{Humbert, M.}, \bibinfo{author}{Brooks, G.}, \bibinfo{author}{Rhamdhani, A.}, \bibinfo{author}{Duffy, A.}, \bibinfo{author}{Pownceby, M.}, \bibinfo{year}{2022}.
\newblock \bibinfo{title}{Mineral {Processing} and {Metal} {Extraction} on the {Lunar} {Surface} - {Challenges} and {Opportunities}}.
\newblock \bibinfo{journal}{Mineral Processing and Extractive Metallurgy Review} \bibinfo{volume}{43}, \bibinfo{pages}{865--891}.
\newblock \DOIprefix\doi{10.1080/08827508.2021.1969390}. \bibinfo{note}{publisher: Taylor \& Francis \_eprint: https://doi.org/10.1080/08827508.2021.1969390}.
%Type = Inproceedings
\bibitem[{Sibille(2020)}]{sibille_state---art_2020}
\bibinfo{author}{Sibille, L.}, \bibinfo{year}{2020}.
\newblock \bibinfo{title}{State-of-the-art and {Systems} {Analysis} of {Emerging} {Technologies} for {Oxygen} and {Metals} {Extraction} from {Lunar} {Regolith}}, \bibinfo{address}{Las Vegas NV}.
\newblock \bibinfo{note}{NTRS Author Affiliations: Southeastern Universities Research Association NTRS Document ID: 20205002032 NTRS Research Center: Kennedy Space Center (KSC)}.
%Type = Article
\bibitem[{Slabic et~al.(2024)Slabic, Technology, Gruener, Rickman, Sibille, Oravec, Edmunson and Keprta}]{slabic_lunar_2024}
\bibinfo{author}{Slabic, A.}, \bibinfo{author}{Technology, J.}, \bibinfo{author}{Gruener, J.E.}, \bibinfo{author}{Rickman, D.L.}, \bibinfo{author}{Sibille, L.}, \bibinfo{author}{Oravec, H.A.}, \bibinfo{author}{Edmunson, J.}, \bibinfo{author}{Keprta, S.}, \bibinfo{year}{2024}.
\newblock \bibinfo{title}{Lunar {Regolith} {Simulant} {User}’s {Guide} {Revision} {A}} .
%Type = Techreport
\bibitem[{{Space Resource Technologies}(2023a)}]{SRT_LHS1_2023}
\bibinfo{author}{{Space Resource Technologies}}, \bibinfo{year}{2023}a.
\newblock \bibinfo{title}{{LHS-1 Lunar Highlands Simulant Fact Sheet}}.
\newblock \bibinfo{type}{Technical Report} \bibinfo{number}{003-01-001-1223}. {Space Resource Technologies}.
\newblock \URLprefix \url{https://spaceresourcetech.com/collections/lunar-simulants/products/lhs-1-lunar-highlands-simulant}. \bibinfo{note}{technical datasheet}.
%Type = Techreport
\bibitem[{{Space Resource Technologies}(2023b)}]{SRT_LHS2_2023}
\bibinfo{author}{{Space Resource Technologies}}, \bibinfo{year}{2023}b.
\newblock \bibinfo{title}{{LHS-2 Lunar Highlands Simulant: Fact Sheet}}.
\newblock \bibinfo{type}{Technical Report} \bibinfo{number}{001-11-001-1223}. {Space Resource Technologies}.
\newblock \URLprefix \url{https://spaceresourcetech.com/collections/lunar-simulants/products/lhs-2-lunar-highlands-simulant}. \bibinfo{note}{technical datasheet}.
%Type = Techreport
\bibitem[{{Space Resource Technologies}(2023c)}]{SRT_LSP2_2023}
\bibinfo{author}{{Space Resource Technologies}}, \bibinfo{year}{2023}c.
\newblock \bibinfo{title}{{LSP-2 Lunar South Pole Simulant: Fact Sheet}}.
\newblock \bibinfo{type}{Technical Report} \bibinfo{number}{001-12-001-1223}. {Space Resource Technologies}.
\newblock \URLprefix \url{https://spaceresourcetech.com/collections/lunar-simulants/products/lsp-2-lunar-south-pole-simulant}. \bibinfo{note}{technical datasheet}.
%Type = Article
\bibitem[{Sun et~al.(2017)Sun, Yi, Shen, Zhang and Ma}]{sun_developing_2017}
\bibinfo{author}{Sun, H.}, \bibinfo{author}{Yi, M.}, \bibinfo{author}{Shen, Z.}, \bibinfo{author}{Zhang, X.}, \bibinfo{author}{Ma, S.}, \bibinfo{year}{2017}.
\newblock \bibinfo{title}{Developing a new controllable lunar dust simulant: {BHLD20}}.
\newblock \bibinfo{journal}{Planetary and Space Science} \bibinfo{volume}{141}, \bibinfo{pages}{17--24}.
\newblock \DOIprefix\doi{10.1016/j.pss.2017.04.010}.
%Type = Article
\bibitem[{Taylor(1972)}]{taylor_composition_1972}
\bibinfo{author}{Taylor, G.J.}, \bibinfo{year}{1972}.
\newblock \bibinfo{title}{The composition of the lunar highlands: evidence from modal and normative plagioclase contents in anorthositic lithic fragments and glasses}.
\newblock \bibinfo{journal}{Earth and Planetary Science Letters} \bibinfo{volume}{16}, \bibinfo{pages}{263--268}.
\newblock \DOIprefix\doi{https://doi.org/10.1016/0012-821X(72)90200-2}.
%Type = Article
\bibitem[{Taylor et~al.(2019)Taylor, Martel, Lucey, Gillis-Davis, Blake and Sarrazin}]{taylor_modal_2019}
\bibinfo{author}{Taylor, G.J.}, \bibinfo{author}{Martel, L.M.}, \bibinfo{author}{Lucey, P.G.}, \bibinfo{author}{Gillis-Davis, J.J.}, \bibinfo{author}{Blake, D.F.}, \bibinfo{author}{Sarrazin, P.}, \bibinfo{year}{2019}.
\newblock \bibinfo{title}{Modal analyses of lunar soils by quantitative {X}-ray diffraction analysis}.
\newblock \bibinfo{journal}{Geochimica et Cosmochimica Acta} \bibinfo{volume}{266}, \bibinfo{pages}{17--28}.
\newblock \DOIprefix\doi{10.1016/j.gca.2019.07.046}.
%Type = Article
\bibitem[{Taylor et~al.(2010)Taylor, Pieters, Patchen, Taylor, Morris, Keller and McKay}]{taylor_mineralogical_2010}
\bibinfo{author}{Taylor, L.A.}, \bibinfo{author}{Pieters, C.}, \bibinfo{author}{Patchen, A.}, \bibinfo{author}{Taylor, D.H.S.}, \bibinfo{author}{Morris, R.V.}, \bibinfo{author}{Keller, L.P.}, \bibinfo{author}{McKay, D.S.}, \bibinfo{year}{2010}.
\newblock \bibinfo{title}{Mineralogical and chemical characterization of lunar highland soils: {Insights} into the space weathering of soils on airless bodies}.
\newblock \bibinfo{journal}{Journal of Geophysical Research: Planets} \bibinfo{volume}{115}.
\newblock \DOIprefix\doi{10.1029/2009JE003427}.
%Type = Misc
\bibitem[{Tewes et~al.(2020)Tewes, Holquist, Bower and Kelsey}]{tewes_isru-derived_2020}
\bibinfo{author}{Tewes, P.}, \bibinfo{author}{Holquist, J.}, \bibinfo{author}{Bower, C.}, \bibinfo{author}{Kelsey, L.}, \bibinfo{year}{2020}.
\newblock \bibinfo{title}{{ISRU}-derived water purification and {Hydrogen} {Oxygen} {Production} ({IHOP}) {Component} {Development}.}
%Type = Article
\bibitem[{Wakita and Schmitt(1970)}]{wakita_lunar_1970}
\bibinfo{author}{Wakita, H.}, \bibinfo{author}{Schmitt, R.A.}, \bibinfo{year}{1970}.
\newblock \bibinfo{title}{Lunar {Anorthosites}: {Rare}-{Earth} and {Other} {Elemental} {Abundances}}.
\newblock \bibinfo{journal}{Science} \bibinfo{volume}{170}, \bibinfo{pages}{969--974}.
\newblock \bibinfo{note}{Publisher: American Association for the Advancement of Science}.
%Type = Article
\bibitem[{Wang et~al.(2025)Wang, Zhang, Wang and Song}]{wang_review_2025}
\bibinfo{author}{Wang, C.}, \bibinfo{author}{Zhang, G.}, \bibinfo{author}{Wang, Y.}, \bibinfo{author}{Song, L.}, \bibinfo{year}{2025}.
\newblock \bibinfo{title}{A {Review} of {Lunar} {Environment} and {In}-{Situ} {Resource} {Utilization} for {Achieving} {Long}-{Term} {Lunar} {Habitation}}.
\newblock \bibinfo{journal}{Galaxies} \bibinfo{volume}{13}, \bibinfo{pages}{103}.
\newblock \DOIprefix\doi{10.3390/galaxies13050103}. \bibinfo{note}{publisher: Multidisciplinary Digital Publishing Institute}.
%Type = Article
\bibitem[{Wang(2022)}]{wang_first_2022}
\bibinfo{author}{Wang, S.J.}, \bibinfo{year}{2022}.
\newblock \bibinfo{title}{First {Location} {And} {Characterization} {Of} {Lunar} {Highland} {Clasts} {In} {Chang}’{E}-5 {Breccias} {Using} {TIMA}-{SEM}-{EPMA}}.
\newblock \bibinfo{journal}{Atomic Spectroscopy} \bibinfo{volume}{43}, \bibinfo{pages}{351--362}.
\newblock \DOIprefix\doi{10.46770/AS.2022.030}.
%Type = Article
\bibitem[{Yin et~al.(2025)Yin, Chen, Fu, Cao, Lu, Liu, Li, Chi, Zeng and Ling}]{yin_petrogenesis_2025}
\bibinfo{author}{Yin, C.}, \bibinfo{author}{Chen, J.}, \bibinfo{author}{Fu, X.}, \bibinfo{author}{Cao, H.}, \bibinfo{author}{Lu, X.}, \bibinfo{author}{Liu, Y.}, \bibinfo{author}{Li, J.}, \bibinfo{author}{Chi, S.}, \bibinfo{author}{Zeng, X.}, \bibinfo{author}{Ling, Z.}, \bibinfo{year}{2025}.
\newblock \bibinfo{title}{Petrogenesis of {Chang}’e-6 {Basalts} and {Implication} for the {Young} {Volcanism} on the {Lunar} {Farside}}.
\newblock \bibinfo{journal}{The Astrophysical Journal Letters} \bibinfo{volume}{981}, \bibinfo{pages}{L2}.
\newblock \DOIprefix\doi{10.3847/2041-8213/adaf20}. \bibinfo{note}{publisher: The American Astronomical Society}.
%Type = Article
\bibitem[{Zhang et~al.(2021)Zhang, Zhang, Zhang, Dong, Deng, Gao, Yang, Xiao, Bai, Liang, Liu, Ma, Zhao, Zhang, Zhang, Song, Yao, Chen, Wang, Zou and Yang}]{zhang_size_2021}
\bibinfo{author}{Zhang, H.}, \bibinfo{author}{Zhang, X.}, \bibinfo{author}{Zhang, G.}, \bibinfo{author}{Dong, K.}, \bibinfo{author}{Deng, X.}, \bibinfo{author}{Gao, X.}, \bibinfo{author}{Yang, Y.}, \bibinfo{author}{Xiao, Y.}, \bibinfo{author}{Bai, X.}, \bibinfo{author}{Liang, K.}, \bibinfo{author}{Liu, Y.}, \bibinfo{author}{Ma, W.}, \bibinfo{author}{Zhao, S.}, \bibinfo{author}{Zhang, C.}, \bibinfo{author}{Zhang, X.}, \bibinfo{author}{Song, J.}, \bibinfo{author}{Yao, W.}, \bibinfo{author}{Chen, H.}, \bibinfo{author}{Wang, W.}, \bibinfo{author}{Zou, Z.}, \bibinfo{author}{Yang, M.}, \bibinfo{year}{2021}.
\newblock \bibinfo{title}{Size, morphology, and composition of lunar samples returned by {Chang}’{E}-5 mission}.
\newblock \bibinfo{journal}{Science China Physics, Mechanics \& Astronomy} \bibinfo{volume}{65}, \bibinfo{pages}{229511}.
\newblock \DOIprefix\doi{10.1007/s11433-021-1818-1}.
%Type = Article
\bibitem[{Zhang et~al.(2025)Zhang, Fa and Jia}]{zhang_provenance_2025}
\bibinfo{author}{Zhang, M.}, \bibinfo{author}{Fa, W.}, \bibinfo{author}{Jia, B.}, \bibinfo{year}{2025}.
\newblock \bibinfo{title}{Provenance and evolution of lunar regolith at the {Chang}’e-6 sampling site}.
\newblock \bibinfo{journal}{Nature Astronomy} \bibinfo{volume}{9}, \bibinfo{pages}{813--823}.
\newblock \DOIprefix\doi{10.1038/s41550-025-02525-7}.
%Type = Article
\bibitem[{Zhong et~al.(2023)Zhong, Low, Zhu, Jiang, Yu, Wang, Zhang, Shang, Long, Yao, Yao, Jiang, Luo, Wang, Yang, Zou and Xiong}]{zhong_situ_2023}
\bibinfo{author}{Zhong, Y.}, \bibinfo{author}{Low, J.}, \bibinfo{author}{Zhu, Q.}, \bibinfo{author}{Jiang, Y.}, \bibinfo{author}{Yu, X.}, \bibinfo{author}{Wang, X.}, \bibinfo{author}{Zhang, F.}, \bibinfo{author}{Shang, W.}, \bibinfo{author}{Long, R.}, \bibinfo{author}{Yao, Y.}, \bibinfo{author}{Yao, W.}, \bibinfo{author}{Jiang, J.}, \bibinfo{author}{Luo, Y.}, \bibinfo{author}{Wang, W.}, \bibinfo{author}{Yang, J.}, \bibinfo{author}{Zou, Z.}, \bibinfo{author}{Xiong, Y.}, \bibinfo{year}{2023}.
\newblock \bibinfo{title}{In situ resource utilization of lunar soil for highly efficient extraterrestrial fuel and oxygen supply}.
\newblock \bibinfo{journal}{National Science Review} \bibinfo{volume}{10}, \bibinfo{pages}{nwac200}.
\newblock \DOIprefix\doi{10.1093/nsr/nwac200}.
%Type = Article
\bibitem[{Zou et~al.(2024)Zou, Wu, Chai, Yang, Ruan and Zhao}]{zou_development_2024}
\bibinfo{author}{Zou, Y.}, \bibinfo{author}{Wu, H.}, \bibinfo{author}{Chai, S.}, \bibinfo{author}{Yang, W.}, \bibinfo{author}{Ruan, R.}, \bibinfo{author}{Zhao, Q.}, \bibinfo{year}{2024}.
\newblock \bibinfo{title}{Development and characterization of the {PolyU}-1 lunar regolith simulant based on {Chang}’e-5 returned samples}.
\newblock \bibinfo{journal}{International Journal of Mining Science and Technology} \bibinfo{volume}{34}, \bibinfo{pages}{1317--1326}.
\newblock \DOIprefix\doi{10.1016/j.ijmst.2024.08.006}.

\end{thebibliography}

\end{document}